\documentclass{article}

\usepackage{microtype}
\usepackage{graphicx}
\usepackage{subcaption}
\usepackage{booktabs}
\usepackage[bottom]{footmisc}
\usepackage[hyphens]{url}
\usepackage[breaklinks]{hyperref}
\hypersetup{colorlinks=true,breaklinks=true}
\usepackage[inkscapelatex=false]{svg}
\usepackage{tabularx}
\usepackage{array}
\usepackage{multirow}

\usepackage[accepted]{icml2024}

\usepackage{amsmath}
\usepackage{amssymb}
\usepackage{mathtools}
\usepackage{amsthm}

\usepackage[capitalize,noabbrev]{cleveref}

\usepackage{paralist}

\theoremstyle{plain}

\theoremstyle{definition}

\theoremstyle{remark}

\usepackage[textsize=tiny]{todonotes}

\icmltitlerunning{Fairness Through Controlled (Un)Awareness in Node Embeddings}

\newcommand{\makecitealias}[1]{\defcitealias{#1}{\citeauthor{#1}, \citeyear{#1}}}
\defcitealias{ai_hleg_high-level_expert_group_on_artificial_intelligence_ethics_2019}{AI HLEG, 2019}
\makecitealias{kordzadeh_algorithmic_2022}
\makecitealias{colquitt_measuring_2015}

\begin{document}

\twocolumn[
\icmltitle{Fairness Through Controlled (Un)Awareness in Node Embeddings}



\icmlsetsymbol{equal}{*}

\begin{icmlauthorlist}
\icmlauthor{Dennis Vetter}{equal,CVAI}
\icmlauthor{Jasper Forth}{equal,TCS}
\icmlauthor{Gemma Roig}{CVAI}
\icmlauthor{Holger Dell}{TCS}
\end{icmlauthorlist}

\icmlaffiliation{CVAI}{Computational Vision and Artificial Intelligence Lab, Goethe University Frankfurt, Frankfurt am Main, Germany}
\icmlaffiliation{TCS}{Theoretical Computer Science Group, Goethe University Frankfurt, Frankfurt am Main, Germany}

\icmlcorrespondingauthor{Dennis Vetter}{vetter@em.uni-frankfurt.de}

\icmlkeywords{machine learning, fairness, social network analysis, graph data, node embedding, representation learning}

\vskip 0.3in
]



\printAffiliationsAndNotice{\icmlEqualContribution} % otherwise use the standard text.

\begin{abstract}
Graph representation learning is central for the application of machine learning (ML) models to complex graphs, such as social networks. Ensuring `fair' representations is essential, due to the societal implications and the use of sensitive personal data. 
In this paper, we demonstrate how the parametrization of the \emph{CrossWalk} algorithm influences the ability to infer a sensitive attributes from node embeddings. By fine-tuning hyperparameters, we show that it is possible to either significantly enhance or obscure the detectability of these attributes. This functionality offers a valuable tool for improving the fairness of ML systems utilizing graph embeddings, making them adaptable to different fairness paradigms.
% % This study evaluates the \emph{CrossWalk} algorithm, designed for fair representations, across multiple social-network datasets.
% % We assess representation quality for both sensitive and non-sensitive attributes using various metrics. 
% % Our results demonstrate \emph{CrossWalk}'s adaptability to different fairness paradigms and highlight that the benefits of these fairness interventions are more pronounced for underrepresented groups.
\end{abstract}

% \begin{abstract}
% Graph representation learning is fundamental for applying machine learning (ML) models to graphs, such as social networks. 
% Ensuring `fair' representations is important due to societal implications and the use of sensitive personal data. 
% This study analyzes the adaptability of the \emph{CrossWalk} algorithm, designed for fair representations, across multiple fairness paradigms. 
% By fine-tuning hyperparameters, we demonstrate how the configuration of the \emph{CrossWalk} algorithm affects the detectability of sensitive attributes from node embeddings. 
% Our results show that it is possible to either significantly increase or reduce the detectability of these attributes. This functionality provides a valuable tool for improving the fairness of ML systems utilizing graph embeddings, making them adaptable to different fairness paradigms and metrics. 
% Moreover, our findings underscore the necessity for responsible use and robust safeguards to prevent potential misuse.
% \end{abstract}

\section{Introduction}
The expanding application of machine learning (ML) systems across various sectors often leads to disparate impact on different subgroups, frequently disadvantaging minorities or marginalized groups \cite{barocas_big_2016, angwin_machine_2016, angwin_facebook_2016, kaplan_measurement_2022}. 
In the context of social networks, these ML systems handle highly personal data and risk negatively impacting lives \cite{angwin_facebook_2016, mello-klein_facebooks_2022, kaplan_measurement_2022, hao_facebooks_2021}.

A common pre-processing step in ML pipelines for large and complex graphs, like social networks, is representation learning, where each node is mapped to a fixed-dimensional real vector that captures relevant structural properties \cite{xu_understanding_2021}. 
Methods for learning node representations include various approaches, with random walks being a common technique exemplified by \emph{DeepWalk} \cite{perozzi_deepwalk_2014} and \emph{node2vec} \cite{grover_node2vec_2016}. 
\emph{DeepWalk} utilizes uniformly random walks to capture network structures while \emph{node2vec} introduces hyperparameters to adapt the walk strategies  for better embeddings in different configurations~\cite{xu_understanding_2021}. 

\citet{khajehnejad_crosswalk_2022} introduced \emph{CrossWalk} to make the node embeddings generated by \emph{node2vec} fairer. In particular, the input vertices are assumed to have \emph{sensitive attributes}, such as \emph{gender} or \emph{ethnicity}, and the random walk is biased to walk towards group boundaries.
% \citet{khajehnejad_crosswalk_2022} introduce \emph{CrossWalk} to enhance fairness in node embeddings using \emph{sensitive attributes}, for example \emph{gender} or \emph{ethnicity}, to bias random walks towards group boundaries. 
This introduces a trade-off between the fairness properties of the node embeddings and the ability for the node embeddings to accurately capture the original graph structure. 
However, the authors do not provide clear guidelines for parameterizing their algorithm and do not evaluate fairness beyond the disparity metric. 
In our work, we address these limitations and evaluate \emph{CrossWalk} using various parametrizations and under different notions of fairness.
% Our study builds on their results by evaluating the algorithm with various parametrizations and under different notions of fairness.
Our contributions are threefold:

\begin{itemize}
    \item We demonstrate how \emph{CrossWalk} can be parameterized to adjust the detectability of sensitive attributes from node embeddings, making it either more difficult or easier to recover these attributes. 
    %control the inference of sensitive attributes from node embeddings.
    
    \item We analyze the impact of fairness improvements on embedding quality by including a non-sensitive control attribute in our investigations.
    
    \item We provide an integrated implementation to apply and evaluate both \emph{CrossWalk} and \emph{node2vec}, enabling easy comparison of their embeddings.
    % \footnote{An open-source Python implementation is available at \href{https://github.com/jasperforth/fairgraphwalk-experiments}{https://github.com/jasperforth/fairgraphwalk-experiments}}
\end{itemize}
% \vspace{-1em}

\section{Related Work}
\subsection{Fairness}
\label{sec:fairness}
In the machine learning literature, the terms `fair' or `just' are often used interchangeably, but they have different meanings in moral philosophy: \emph{Fairness} is subjective and dependent on personal perception, whereas \emph{justice} is objective, relying on external standards such as laws or other rules. 
This distinction is relevant when defining ML `fairness metrics', which do not measure subjective fairness, but align more with the philosophical concept of justice~\cite{kordzadeh_algorithmic_2022, colquitt_measuring_2015, goldman_justice_2015}. 

\emph{Fairness} in ML typically involves grouping the population based on `sensitive attributes' like ethnicity, gender, or age \cite{mehrabi_survey_2022, barocas_big_2016, kuppler_distributive_2021}. 
In our work, we focus on two concepts for fair ML: \emph{procedural justice} and \emph{distributive justice}. Procedural justice concerns just processes \cite{rawls_theory_1999, miller_justice_2023}, aiming to prevent ML systems from using sensitive attributes~\cite{kusner_counterfactual_2017, grgic-hlaca_case_2016}. This approach faces challenges: 
\begin{inparaenum}[(1)]
    \item the system should not infer group membership from non-sensitive attributes, and
    \item excluding data might worsen results~\cite{barocas_big_2016, grgic-hlaca_case_2016, kozyrkow_ai_2023}.
\end{inparaenum}

In contrast, distributive justice focuses on the just allocation of outcomes~\cite{rawls_theory_1999, miller_justice_2023}, ensuring similar performance across sub-groups~\cite{mehrabi_survey_2022, kordzadeh_algorithmic_2022, kuppler_distributive_2021}. 
The choice between procedural and distributive justice depends on the use-case, highlighting the need for a holistic approach in ethical applications of ML~\citepalias{kordzadeh_algorithmic_2022, colquitt_measuring_2015, ai_hleg_high-level_expert_group_on_artificial_intelligence_ethics_2019}.

\subsection{Learning Graph Representations}
In the context of our work, the goal of representation learning is to create an \emph{embedding}, mapping graph nodes to fixed-dimensional real vectors while preserving node relations and features ~\cite{perozzi_deepwalk_2014, grover_node2vec_2016}. This enables the application of standard statistical and ML approaches. Node similarity can be defined by \emph{homophily} -- similar neighbors, or \emph{structural equivalence} -- similar neighborhood structures. 

A common method for learning embeddings uses the \emph{Skip-Gram} model~\cite{mikolov_efficient_2013}, often applied in natural language processing. 
In this analogy, nodes are `words' and walks are `sentences' \cite{perozzi_deepwalk_2014, tang_line_2015, grover_node2vec_2016}. Under this model, the strategy that produces the walks is important.

\emph{DeepWalk} performs random walks by uniformly sampling nodes and their neighbors \cite{perozzi_deepwalk_2014}.
\emph{node2vec} generalizes the sampling process by using edge weights as transition probabilities and introducing two hyperparameters $p$ and $q$. The parameter $p$ controls exploration away from the walk's root, while $q$ balances between homophily and structural equivalence \cite{grover_node2vec_2016}. 
%\emph{FairWalk} \cite{rahman_fairwalk_2019} extends node2vec towards producing fairer embeddings, by introducing a different bias in the random walk, focused on fairness. Instead of randomly selecting from all neighbors of a node, in FairWalk, neighbors are partitioned into groups based on a sensitive attribute. The walk then selects a group uniformly at random before choosing a node within that group. 

\emph{CrossWalk} \cite{khajehnejad_crosswalk_2022} employs a fairness intervention before learning representations with \emph{node2vec}. It groups nodes by a sensitive attribute and adjusts transition probabilities to favor crossing group boundaries, reducing the distance in embedding space between nodes from different groups.
CrossWalk uses two hyperparameters, $\alpha$ and~$\beta$.\footnote{\citet{khajehnejad_crosswalk_2022} use the names $\alpha$ and $p$, we changed~$p$ to~$\beta$ to avoid confusion with \emph{node2vec}'s $p$.}
The parameter $\alpha$ controls the likelihood of crossing group boundaries, while $\beta$ encourages visits near group boundaries.

\section{Experimental setup}
\subsection{Datasets}
\label{sec:datasets}
Obtaining diverse graph datasets with multiple attributes per node proved challenging. To adress this, we utilized demographic-rich data from the Pokec social network and created multiple controlled subgraph datasets to evaluate \emph{CrossWalk} in different settings.

The base of our datasets is the Pokec social network dataset, a collection of anonymized user data, including attributes like age, place of living, and connections with other users \cite{takac_data_2012}, sourced from SNAP \cite{leskovec_snap_2014}. The total graph comprises over 1.6 million nodes and over 30 million edges \cite{takac_data_2012}. 

In our experiments, we focus on the attributes \emph{age} and \emph{location}, and we select only nodes where these attributes are available. Preliminary analysis revealed that nodes with similar attributes tend to be more densely connected.
 
We define three categories of subgraphs: 
\begin{inparaenum}[(1)]
    \item \emph{distinct}: small towns characterized by few connections between each other, due to their geographical separation; 
    \item \emph{semi-distinct}: clusters of well-connected small towns in geographic proximity, with few connections between geographically distant groups; and 
    \item \emph{mixed}: adjacent city districts within the same urban area.
\end{inparaenum}

\begin{table}[ht]
    \centering
    \caption{Summary of the experiment data. Each category consists of three subgraphs with varying numbers of nodes, edges, and location groups.}
    \label{tab:experiment_data}
        % \footnotesize
        % \renewcommand\arraystretch{1.5}
    \begin{tabularx}{0.472\textwidth}{lcrrcc}
      \toprule
          & \textbf{Sub-} & \textbf{\#Nodes} & \textbf{\#Edges} & \multicolumn{2}{c}{\textbf{\#Groups}} \\ 
          &       \textbf{Graph}            &                  &                  & \textbf{Location}  & \textbf{Age} \\
      \midrule
          \multirow{3}{*}{\rotatebox[origin=c]{90}{Distinct}} 
          & 0 & 2,504  & 10,685  & 3 & 3 \\
          & 1 & 20,927 & 127,329 & 4 & 3 \\
          & 2 & 24,228 & 145,488 & 8 & 3 \\
      \midrule
          \multirow{3}{*}{\rotatebox[origin=c]{90}{Semi}} 
          & 0 & 9,240  & 33,961  & 4 & 3 \\
          & 1 & 32,640 & 183,478 & 12 & 3 \\
          & 2 & 35,086 & 232,304 & 14 & 3 \\
      \midrule
          \multirow{3}{*}{\rotatebox[origin=c]{90}{Mixed}} 
          & 0 & 15,677 & 26,339  & 5 & 3 \\
          & 1 & 32,973 & 180,788 & 8 & 3 \\
          & 2 & 49,222 & 138,405 & 12 & 3 \\
          \bottomrule
    \end{tabularx}
    % \vspace{0.1cm}
\end{table}

For each category, we compile three different datasets with manually selected locations based on geographical proximity. This approach allowed us to control the number of connections between nodes from different locations, creating `easier' (distinct category) and `harder' (mixed category) node classification problems.
We observed that locations are internally well-connected, and geographically close locations share more connections than geographically distant ones.
An example of this can be seen in Figure~\ref{fig:semi-distinct-graph} in the appendix,  which illustrates a graph from the \emph{semi-distinc} category. This pattern of internal connectivity and proximity-based connections holds across all other datasets.

In addition, we categorize the age attribute into three groups: 
\begin{inparaenum}[(a)]
    \item 16-18 years, 
    \item 19-21 years, and 
    \item 22 years and older, 
\end{inparaenum}
to obtain differing group sizes and approximately equal levels of interconnectedness that are more stable across the different subgraph categories.
With this, we can analyze how the performance of the \emph{CrossWalk} algorithm changes under different pre-conditions, and how the generated representations perform with respect to a non-sensitive attribute.
\cref{tab:experiment_data} summarizes the different controlled subgraph selections used in our experiments.
% \cref{tab:experiment_data} provides an overview of the different controlled subgraph selections we use for our experiments. Each subgraph has a varying number of nodes, edges and number of classes for the location attribute, while all of them share the same number of classes for the age attribute. This allows us to have different levels of complexity in predicting the `location' attribute, and approximately similar complexity for predicting the `age' attribute.

\subsection{Experimental Pipeline}
\label{sec:setup-pipeline}
Our experimental pipeline is structured as follows:
% \vspace{-1em}
\begin{enumerate}%[itemsep=2mm]
    \item \emph{Dataset selection:} We choose a subgraph from one of the three categories: \emph{distinct}, \emph{semi-distinct}, or \emph{mixed}. 
    \item \emph{CrossWalk biasing:} We select parameters ($\alpha$ and $\beta$) and a sensitive attribute (`location' or `age') to bias the transition probabilities within the graph. 
    \item \emph{node2vec embedding:} We run the \emph{node2vec} algorithm with a set of parameters ($p$ and $q$) to generate node embeddings through a collection of random walks. 
    \item \emph{Attribute prediction:} From the resulting node embeddings, we predict either the sensitive attribute or a control attribute (`location' or `age') using the Label Propagation algorithm \cite{zhu_learning_2002} with 50\% of nodes labeled. We employed 25-fold cross-validation for robust performance estimation.
\end{enumerate}

% Note that this experimental pipeline can easily be adapted to use different algorithms to bias the transition probabilities in the graph, where we currently use \emph{CrossWalk}, or to sample the random walks, where we currently use node2vec.
% To ensure consistency, the attributes `location' and `age' are used once as the sensitive and the control attribute in each case. 
% \cref{tab:experiment_data} provides an overview of the groups and graph sizes for each category. 

The effect of the hyperparameters of \emph{node2vec} ($p,q$) and \emph{CrossWalk} ($\alpha, \beta$) is summarized in \cref{tab:parameter_summary}. 
We evaluated all combinations of the hyperparameters with the following values:
\begin{itemize}
    \item[--] $p$, $q$: 0.1, 0.5, 1, 5.0, 10.0
    \item[--] $\alpha$: 0.01, 0.25, 0.5, 0.75, 0.99
    \item[--] $\beta$: 1, 2, 3, 5, 8, 11, 15
\end{itemize}
% All combinations of these hyperparameters were evaluated. \cref{tab:parameter_summary} summarizes the impact these hyperparameters.
\begin{table}[htbp!]
\centering
\caption{Effect of the hyperparameters in \emph{node2vec} ($p,q$) and \emph{CrossWalk} ($\alpha, \beta$).}
\begin{tabular}{cp{5.98cm}}
\hline
\textbf{Param.} & \textbf{Effect} \\ \hline
\multirow{3}{*}{$p$} & Controls walk distance from root. \\ 
                     & $p < \min(q,1)$ for local revisits \\ 
                     & $p > \max(q,1)$ for exploration \\ \hline
\multirow{3}{*}{$q$} & Controls notion of similarity. \\ 
                     & $q < 1$ for DFS style (homophily) \\ 
                     & $q > 1$ for BFS style (structural equivalence) \\ \hline
\multirow{3}{*}{$\alpha$} & Controls group boundary crossing. \\ 
                          & Higher values make crossing group boundaries more likely. \\ \hline
\multirow{3}{*}{$\beta$} & Encourages walks towards nodes on group boundaries. Higher values correspond to stronger encouragement. \\ \hline
\end{tabular}
\label{tab:parameter_summary}
\end{table}

We further alternate between using \emph{location} and \emph{age} as the sensitive attributes to assess the adaptability of \emph{Crosswalks}'s fairness intervention. 
The source code to create controlled subgraph datasets, to bias the transition probabilities, to generate the random walks, and to perform the experiments is available under an open-source license.\footnote{\href{https://github.com/jasperforth/fairgraphwalk-experiments}{https://github.com/jasperforth/fairgraphwalk-experiments}}

\begin{figure*}[ht]
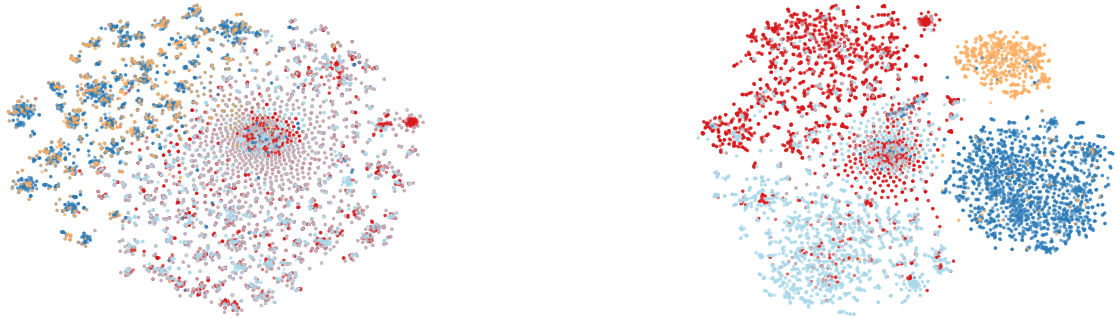

    \centering
    \begin{subfigure}[b]{0.46\textwidth}
        \centering
        \includesvg[width=0.80\textwidth]{Figures/tsne/pokec_semi_0__alpha_0.99_exponent_15_p_1_q_1_other_AGE.svg}
        \caption{
        The \emph{low awareness} configuration improves procedural justice, as inferring the sensitive attribute is more difficult.
        }
    \end{subfigure}
    \hfill
    \begin{subfigure}[b]{0.46\textwidth}
        \centering
        \includesvg[width=0.80\textwidth]{Figures/tsne/pokec_semi_0__alpha_0.01_exponent_1_p_1_q_1_other_AGE.svg}
        \caption{The \emph{high awareness} configuration improves distributive justice, as inferring the sensitive attribute is easier for all groups.}
        \label{fig:embeddings-distributive}
    \end{subfigure}
    \vspace{-0.5em}
    \caption{Impact of \emph{low awareness} and \emph{high awareness}  configurations of the \emph{CrossWalk} algorithm on node embeddings for a subgraph from the \emph{semi-distinct} category, with sensitive attribute `location'.
    The embeddings are visualized using t-SNE for dimensionality reduction, and the node colors correspond to sensitive attribute class.}
    \label{fig:embeddings}
\end{figure*}

\subsection{Evaluation Method}
\label{sec:evaluation-method}
Our evaluation focuses on three key aspects: 
\begin{inparaenum}[(i)]
    \item awareness, 
    \item disparity, and 
    \item classification performance. 
\end{inparaenum}
The baseline for our analysis consists of embeddings created by \emph{node2vec} without \emph{CrossWalk} biasing. Since \emph{node2vec} embeddings are produced without using node attributes, this baseline helps us analyze the effect of \emph{CrossWalk}'s fairness intervention.

Let $G=(V,E)$ be a graph where each node $v \in V$ is associated with a sensitive attribute and a control attribute, and the nodes are partitioned into $C$ distinct groups $1, \ldots, C$ based on the sensitive attribute. 
For each group $i \in \{ 1, \ldots, C\}$ defined by the sensitive attribute, we write $Q_i\in[0,1]$ for the ability to predict it, measured as F1 score of Label Propagation, and let $Q^*_i\in[0,1]$ be the corresponding prediction F1 score for the control attribute, based on the same sensitive attribute groups. 
% to predict the sensitive attribute from the embeddings for group $i \in \{ 1, \ldots, C\}$ and $Q^*_i$ for the non-sensitive attribute. 
% We define $\mathrm{awareness} := \mathrm{max}(Q_i:i\in {1, \ldots, C})$ as the ability to infer group membership from the embedding. Lower awareness indicates that it is more difficult to infer the group membership from the embeddings, which improves fairness under procedural justice.
% We define:
% \begin{itemize}
%     \item \textbf{Awareness ($\mathrm{awareness}$):} $\mathrm{max}(Q_i \colon i \in {1, \ldots, C})$ - 
%     This measures the ability to infer group membership from the embeddings. 
%     Lower awareness indicates improved procedural justice.
%     \item \textbf{Disparity ($\mathrm{disparity}$):} $\mathrm{var}(Q_{i}\colon i \in {1, \ldots, C})$ - 
%     This compares performance across different groups. 
%     Lower disparity indicates improved distributive justice.
%     \item \textbf{Performance ($\mathrm{performance}_s$):} $\mathrm{mean}(Q_i \colon i \in {1, \ldots, C})$ for the sensitive attribute and $\mathrm{performance_c}$ for the non-sensitive attribute. 
%     High performance for the control attribute ($\mathrm{performance}_c$) is desirable as it indicates that the \emph{CrossWalk} intervention does not degrade the quality of embeddings for non-sensitive attributes.
% \end{itemize}
% These measures allow us to comprehensively evaluate the impact of the \emph{CrossWalk} fairness intervention on both sensitive and non-sensitive attributes.
We use the following metrics in our evaluation:
\vspace{-0.75em}
    \paragraph{Awareness.} Defined as $\mathrm{max}(Q_i \colon i \in \{1, \ldots, C\})$, this metric captures the ability to infer group membership from the embeddings. Lower \emph{awareness} indicates better procedural justice.
\vspace{-0.75em}    
    \paragraph{Disparity.} Following \citet{khajehnejad_crosswalk_2022}, we measure \emph{disparity} as $\mathrm{var}(Q_{i} \colon i \in \{1, \ldots, C\})$. This metric compares the classification F1 score across different groups, with lower {disparity} indicating better distributive justice.
\vspace{-0.75em}    
    \paragraph{Performance.} We use \emph{performance} as $\mathrm{mean}(Q^*_i \colon i \in \{1, \ldots, C\})$, to measure the ability of correctly predicting the control attribute across sensitive attribute groups.
    % for the control attribute, $\emph{performance}_c$ is defined analogously as the mean of the $Q_i^*$. 
    If \emph{performance} stays high regardless of whether the \emph{CrossWalk} intervention was used before applying \emph{node2vec}, then applying \emph{CrossWalk} does not degrade the quality of embeddings for non-sensitive attributes. 

\section{Results}
\label{sec:results}
This section presents the impact of \emph{node2vec} and \emph{CrossWalk} parametrizations on our evaluation metrics: awareness, disparity, and classification performance.

\begin{figure}[h]
    \centering
    \includesvg[width=0.9\linewidth]{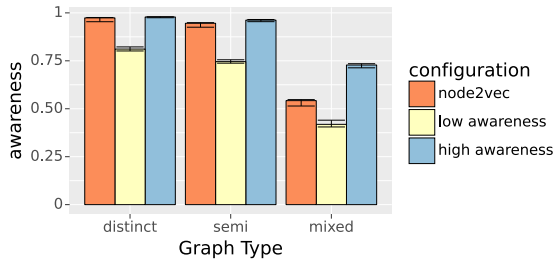}
    \caption{Mean awareness over subgraphs for sensitive attribute `location'. 
    % \emph{CrossWalk}'s impact on awareness for sensitive attribute `location'. 
    Error bars show range of values over all \emph{node2vec} parametrisations. \emph{Low} and \emph{high awareness} configurations can consistently adapt awareness below or above the \emph{node2vec} baseline.}
    \label{fig:res-awareness}
\end{figure}

Our experiments show that the \emph{node2vec} parametrization alone does not significantly influence our evaluation metrics. This can be seen in \cref{fig:res-awareness}, where the error bars indicate the value range across all evaluated combinations of \emph{node2vec} hyperparameters. 
% However, there is one notable exception. Configurations with small values for $p$ and large values for $q$ lead to a decrease in all our observed metrics. This combination corresponds to random walks that emphasize structural equivalence, which can be ineffective when homophily is more important. 
% In social networks, homophily suggests that nodes with similar labels are more likely to connect, whereas structural equivalence focuses on nodes having similar neighborhood structures.
% Thus, overly sampling for structural equivalence results in low-performing embeddings for tasks relying in homophily.
%
In contrast, the \emph{CrossWalk} parameters have a much higher impact on the representations. We identify two main configurations: \emph{low awareness} and \emph{high awareness}, illustrated by \cref{fig:embeddings}.

\emph{Low awareness} is achieved through high values for the hyperparameters $\alpha$ and $\beta$. In our experiments this corresponds to $\alpha = 0.99$ and $\beta = 15$. With this configuration, the random walks are biased towards visiting nodes at group boundaries and frequently crossing group boundaries. This leads to embeddings where the groups are much more different to separate.
% e resulting embeddings are not sensitive to the group membership of the nodes. 
This intermixing of embeddings is more effective when the groups are well interconnected. 

\cref{fig:embeddings} shows the resulting embeddings for the same dataset from the \emph{semi-distinct} category under the two different configurations. 
\cref{fig:res-awareness} indicates that these results are consistent across the three graph categories. 
In the \emph{low awareness} configuration depicted in \cref{fig:embeddings}(a), embeddings from geographically close locations, such as yellow and dark blue, are much more intermixed than the embeddings of nodes from geographically distant locations, such as yellow and red. 
Another important observation is that while \emph{CrossWalk} is able to mix the embeddings from the geographically close locations, in none of our evaluated configurations were the embeddings from geographically distant locations intermixed.
A possible cause could be that there are simply not enough interconnections between the groups that allow biased random walks to cross these specific group borders.

\emph{High awareness} is achieved through low values for the hyperparameters of $\alpha$ and $\beta$. In our experiments, this corresponds to $\alpha = 0.01$ and $\beta=1$. This configuration discourages the crossing of boundaries and biases the random walks towards staying in the same group as the start node. This results in similar embeddings for nodes from the same group and different embeddings for nodes from different groups, as seen in \cref{fig:embeddings}(b). 

In addition, it can be seen that the \emph{CrossWalk} intervention makes it possible to increase awareness above or below the \emph{node2vec} baseline.

\begin{figure}[h]
    \centering
    \includesvg[width=0.93\linewidth]{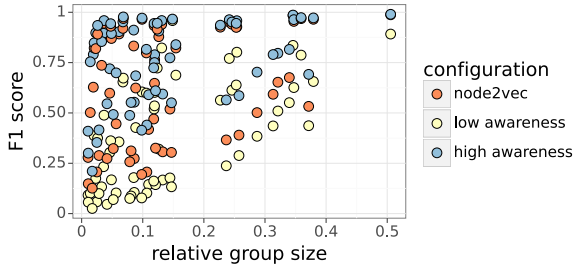}
    \caption{Classification F1 score for sensitive attribute `location' by relative group size of groups in all subgraphs. \emph{Low awareness} makes classification of small groups virtually impossible, differences in F1 score between \emph{low} and \emph{high awareness} are more pronounced for smaller groups.}
    \label{fig:unequal-change}
\end{figure}

\begin{figure}[h]
    \centering
    \includesvg[width=0.93\linewidth]{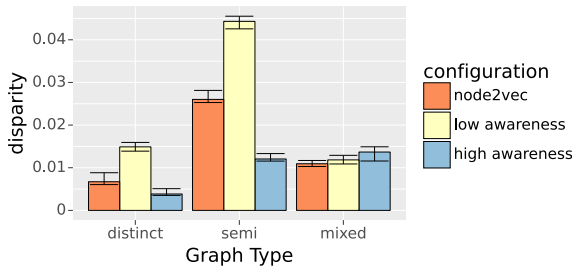}
    \caption{\emph{CrossWalk}'s impact on disparity of sensitive attribute `location'. The \emph{low awareness} configuration tends to lead to higher disparity, indicating an unequal impact on the different groups.
    % varying based on the level of intermixing among the groups. 
    % indicating that the impact on classification score differs between groups changes in awareness for different groups.
    }
    \label{fig:res-disparity}
\end{figure}

We also found that the \emph{low awareness} configuration tends to increase the \emph{disparity} metric, which is highlighted in \cref{fig:res-disparity}. 
The combination of reduced awareness and increased disparity between the \emph{low awareness} and \emph{high awareness} configurations suggests that the changes in classification score are not equally distributed.
As seen in \cref{fig:unequal-change}, this difference tends to be more pronounced for smaller groups.
% Closer investigation indicates that this is caused by unequally distributed changes in classification performance. 
% As seen in \cref{fig:unequal-change}, the difference between classification performance in the \emph{high} and \emph{low awareness} configurations is especially pronounced for smaller groups. 

In the \emph{high awareness} configuration, the classification F1 score across groups becomes more equal, leading to reduced \emph{disparity} (see Figure 3) and improved distributive justice. Conversely, in the \emph{low awareness} configuration, it can become virtually impossible to correctly classify members of underrepresented groups, thereby improving procedural justice.
This unequal impact on classification F1 score can itself be seen as just under the criterion of prioritarianism, where more resources should be allocated to the individuals or groups that are worse off \cite{kuppler_distributive_2021, parfit_equality_2001}. 

% \begin{figure*}
%     \centering
%     \includesvg[width=0.93\linewidth]{Figures/workshop/acc_by_group_location.svg}
%     \caption{}
%     \label{fig:acc-group}
% \end{figure*}

\begin{figure}[h]
    \centering
    \includesvg[width=0.93\linewidth]{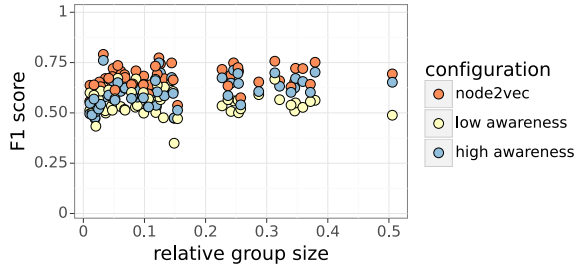}
    \caption{Classification F1 score for the control attribute `age' by relative group size of groups in all subgraphs.
    Influencing embeddings for lower or higher awareness of the sensitive attribute makes classifying the control attribute more difficult.}
    \label{fig:res-performance}
\end{figure}

Finally, we investigate how the \emph{CrossWalk} fairness intervention impacts the classification performance of the non-sensitive attribute. As seen in \cref{fig:res-performance}, applying both \emph{low awareness} and \emph{high awareness} \emph{CrossWalk} configurations leads to decreases in classification performance for the non-sensitive attribute. 
Furthermore, across all of our conducted experiments, no configuration of \emph{CrossWalk} hyperparameters could improve \emph{performance} over the \emph{node2vec} baseline. 
This leads to the conclusion that \emph{CrossWalk's} introduces a trade-off between improving fairness for the sensitive attribute and accurately capturing the graph structure.
% Our data suggests that the more the \emph{CrossWalk} fairness intervention improves information on the sensitive attribute in the embeddings, the more it decreases information on the non-sensitive attribute.

\section{Conclusion}
Our findings reveal that \emph{CrossWalk} can be strategically parameterized to significantly alter the ability to infer sensitive attributes. 
This alteration manifests in two principal ways:
\begin{inparaenum}[(1)]
    \item by enhancing procedural justice through the obfuscation of the sensitive attributes, and 
    \item by promoting distributive justice via reducing disparities in the classification performance of sensitive attributes. 
\end{inparaenum}

We demonstrate that these goals are readily achievable: high hyperparameter lead to low awareness of the sensitive attribute, enhancing procedural justice, while low values lead to high awareness of the sensitive attribute, improving distributive justice.
Moreover, our results indicate that embeddings generated with \emph{CrossWalk} bias generally outperform those created with node2vec alone regarding these notions of justice. Additionaly, underrepresented groups benefit more from the \emph{CrossWalk} fairness intervention than majority groups. 

However, this approach does present a notable trade-off: while it elevates fairness concerning sensitive attributes, it can diminish performance and fairness for other attributes. An important limitation of the \emph{CrossWalk} algorithm is its dependency on prior knowledge of sensitive attributes for biasing.
This makes it a powerful tool for deliberately incorporating such information into node embeddings but also presents the potential for misuse. For instance, rather than diminishing the recognition of sensitive attributes, \emph{CrossWalk} could be misused to more effectively identify minority groups, possibly resulting in adverse outcomes for these groups.

Overall, the \emph{CrossWalk} algorithm emerges as a potent tool for learning fairer node embeddings. However, it is important to acknowledge that no single configuration leads to ideal results in every scenario. With the parametrization guidelines presented in this work, \emph{CrossWalk} can be effectively tailored to provide use-case dependent solutions, adeptly navigating the various trade-offs it presents.

\section*{Impact Statement}
In this paper, we demonstrate how the parametrization of the \emph{CrossWalk} algorithm influences the ability to infer a sensitive attributes from node embeddings. By fine-tuning hyperparameters, we show that it is possible to either significantly enhance or obscure the detectability of these attributes. This functionality offers a valuable tool for improving the fairness of ML systems that rely on graph embeddings, making them adaptable to different fairness paradigms.

However, this technology also raises concerns about potential misuse. Enhanced detection of underrepresented groups, such as those based on religion or sexuality, could be exploited, leading to adverse consequences for these individuals. The \emph{CrossWalk} fairness intervention requires the presence of sensitive attribute values in the dataset, implying that malicious actors already have access to information that could harm underrepresented groups. While the ability to manipulate the prominence of sensitive attributes in embeddings does not fundamentally alter this risk, it underscores the need for responsible use and robust safeguards to prevent misuse.

\bibliography{references.bib}
\bibliographystyle{icml2024}

%%%%%%%%%%%%%%%%%%%%%%%%%%%%%%%%%%%%%%%%%%%%%%%%%%%%%%%%%%%%%%%%%%%%%%%%%%%%%%%
%%%%%%%%%%%%%%%%%%%%%%%%%%%%%%%%%%%%%%%%%%%%%%%%%%%%%%%%%%%%%%%%%%%%%%%%%%%%%%%
% APPENDIX
%%%%%%%%%%%%%%%%%%%%%%%%%%%%%%%%%%%%%%%%%%%%%%%%%%%%%%%%%%%%%%%%%%%%%%%%%%%%%%%
%%%%%%%%%%%%%%%%%%%%%%%%%%%%%%%%%%%%%%%%%%%%%%%%%%%%%%%%%%%%%%%%%%%%%%%%%%%%%%%
\newpage
\appendix
\onecolumn
\section{Geographical- and Force Directed Layout of Subgraph `semi-distinct 0'}
\begin{figure}[h]
    \centering
    \begin{subfigure}[c]{0.48\textwidth}
        \centering
        \includegraphics[width=\textwidth]{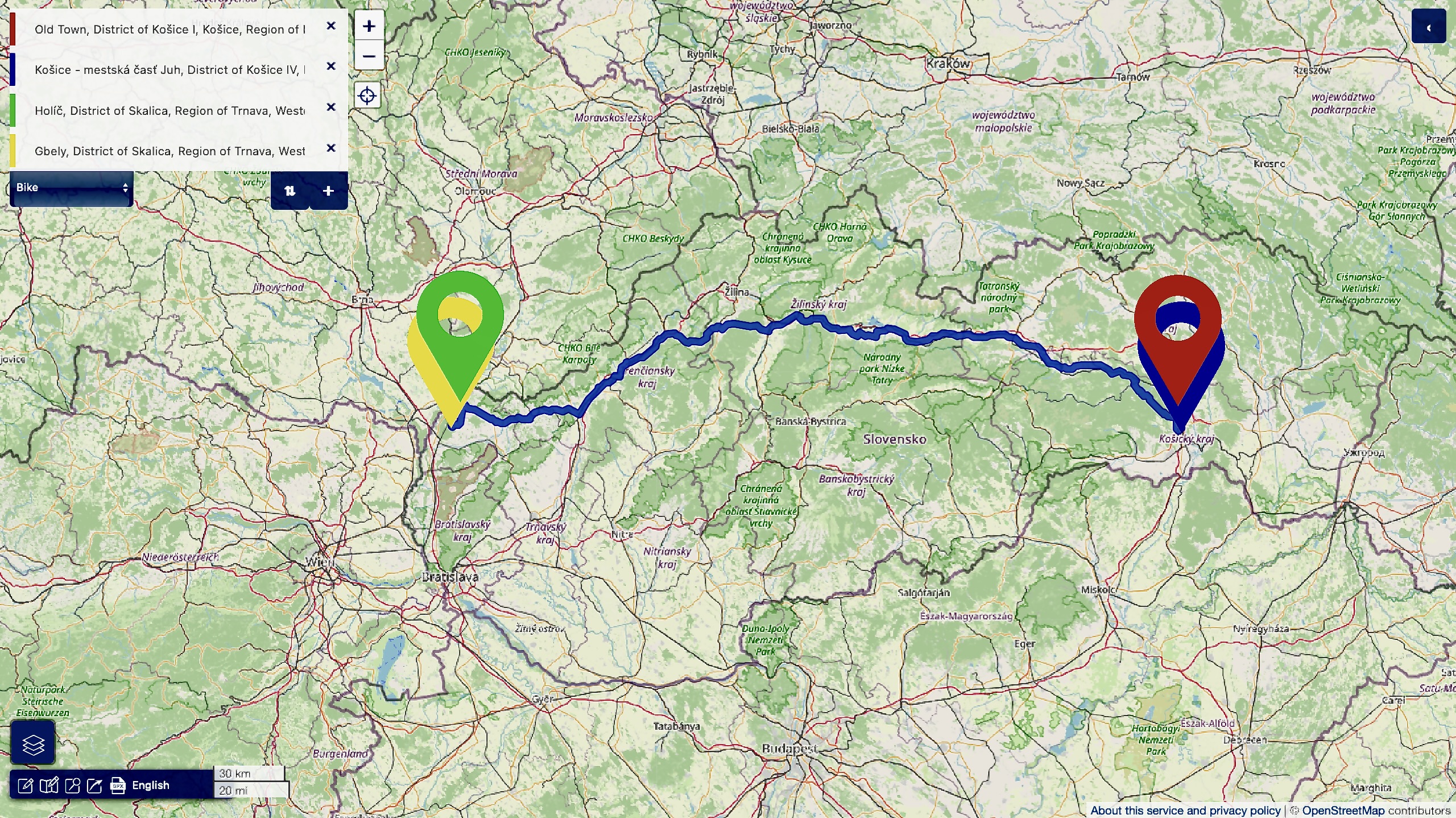}
        \caption{Geographical layout: Displays four towns, each marked with a unique color, representing their geographical locations as foundation for the graph selection.}
        \label{fig:semi-distinct-graph map}
    \end{subfigure}
    \hfill
    \begin{subfigure}[c]{0.48\textwidth}
        \centering
        \includegraphics[width=\textwidth]{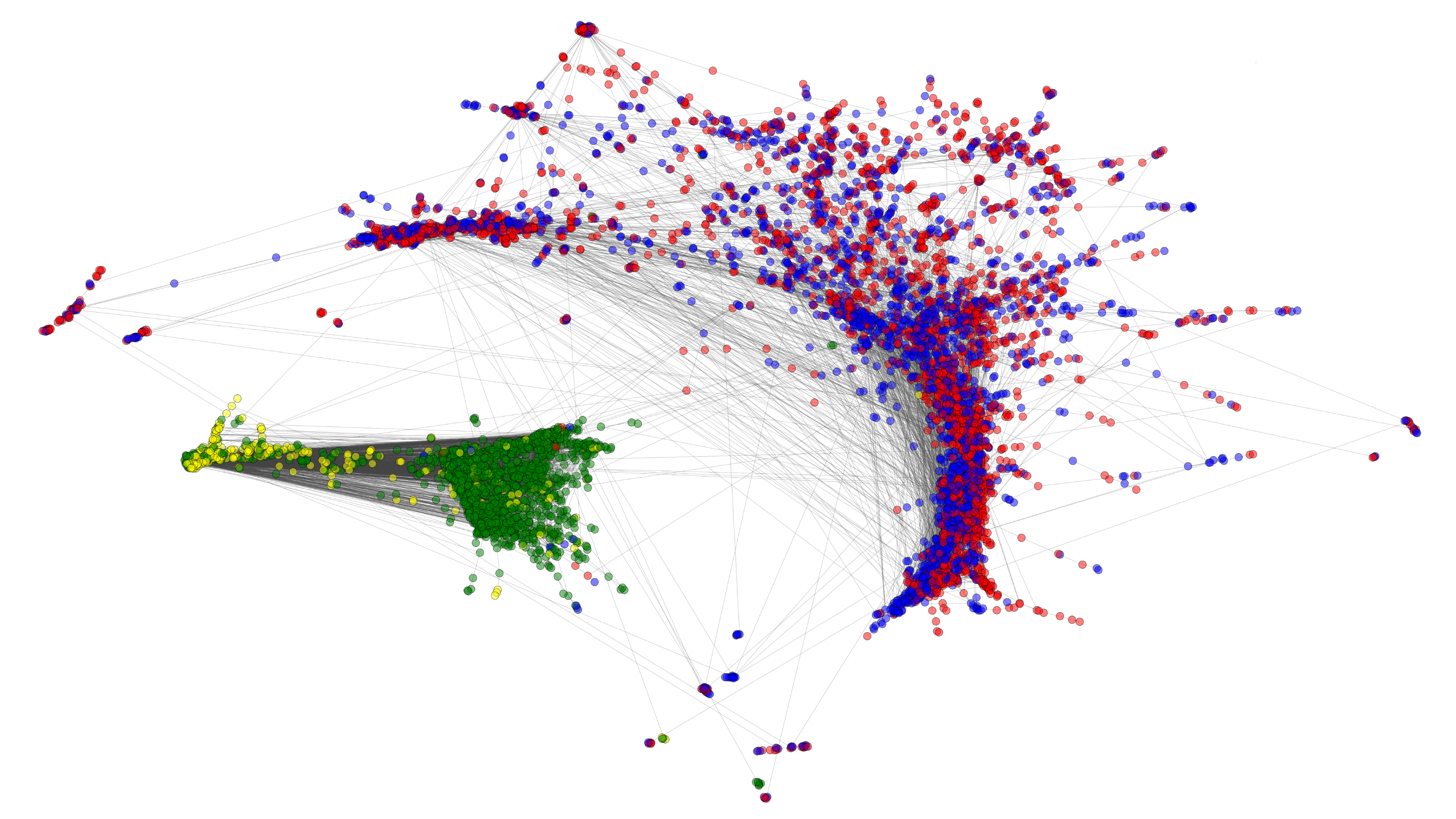}
        \caption{Graph visualization: Illustrates the social network of individuals from these towns, with each node representing an individual and its color, representing the location group 
        corresponding to the respective town on the map.}
        \label{fig:semi-distinct-graph embedding}
    \end{subfigure}
    \caption{Visualization of a semi-distinct graph from our dataset. The left panel shows the geographical positioning of four towns, while the right panel shows the social network derived from these locations. This figure demonstrated the correlation between geographical proximity and network connectivity, where individuals from from geographically closer towns show higher interconnectivity.}
    \label{fig:semi-distinct-graph}
\end{figure}
% You can have as much text here as you want. The main body must be at most $8$ pages long.
% For the final version, one more page can be added.
% If you want, you can use an appendix like this one.  

% The $\mathtt{\backslash onecolumn}$ command above can be kept in place if you prefer a one-column appendix, or can be removed if you prefer a two-column appendix.  Apart from this possible change, the style (font size, spacing, margins, page numbering, etc.) should be kept the same as the main body.
%%%%%%%%%%%%%%%%%%%%%%%%%%%%%%%%%%%%%%%%%%%%%%%%%%%%%%%%%%%%%%%%%%%%%%%%%%%%%%%
%%%%%%%%%%%%%%%%%%%%%%%%%%%%%%%%%%%%%%%%%%%%%%%%%%%%%%%%%%%%%%%%%%%%%%%%%%%%%%%

\end{document}